# First Integrated Implosion Experiment of Three-Axis Cylindrical Hohlraum at the SGIII Laser Facility

Longyu Kuang,[1,2] Hang Li,[1,2] Shaoen Jiang,[1,*] Longfei Jing,[1] Jianhua Zheng,[1] Liling Li,[1] Zhiwei Lin,[1] Lu Zhang,[1] Yulong Li,[1] Xiangming Liu,[1] Xiaoshi Peng,[1] Qi Tang,[1] Xiayu Zhan,[1] Zhurong Cao,[1] Qiangqiang Wang,[1] Bo Deng,[1] Keli Deng,[1] Lifei Hou,[1] Huabing Du,[1] Wei Jiang,[1] Zhongjing Chen,[1] Dong Yang,[1] Feng Wang,[1] Jiamin Yang,[1] Lin Gao,[1] Haijun Zhang,[1] Juan Zhang,[1] Jun Xie,[1] Guanghui Yuan,[1] Zhibing He,[1] Wei Zhou,[1] Yuancheng Wang,[1] Xiaoxia Huang,[1] Xu Chen,[1] Caibo Deng,[1] Baohan Zhang,[1] Jian Zheng,[2] Yongkun Ding,[2,3,†]

[1]*Laser Fusion Research Center, China Academy of Engineering Physics, P.O.Box 919-986, Mianyang 621900, China*

[2]*CAS Key Laboratory of Basic Plasma Physics and Department of Modern Physics, University of Science and Technology of China, Hefei 230026, China*

[3]*Institute of Applied Physics and Computational Mathematics, Beijing 100088, China*

Abstract: The first integrated implosion experiment of three-axis cylindrical hohlraum (TACH) was accomplished at the SGIII laser facility. 24 laser beams of the SGIII laser facility were carefully chosen and quasi-symmetrically injected into the TACH, in which a highly symmetric radiation filed was generated with a peak radiation temperature of ~190eV. Driven by the radiation field, the neutron yield of a deuterium gas filled capsule reached ~$1\times10^9$, and the corresponding yield over clean (YOC) was ~40% for a convergence ratio (Cr) of ~17. The X-ray self-emission image of imploded capsule cores was nearly round, and the backscatter fraction of laser beams was less than 1.25%. This experiment preliminarily demonstrated the major performance of TACH, such as the robustness of symmetry, and a laser plasma instability (LPI) behavior similar to that of the outer ring of traditional cylindrical hohlraum.

Introduction: Hohlraum is an important part in indirect-drive Inertial Confinement Fusion (ICF). Laser beams are injected into a high-Z hohlraum through the laser entrance holes (LEHs) and converted into quasi-blackbody radiation. Then the radiation drives a low-Z capsule placed at the center of the hohlraum to the ignition condition [1,2] with a Cr of 30~40. To achieve this goal, the radiation asymmetry should be less than 1%. To obtain the desired radiation symmetry, hohlraum structure and laser beams arrangement should be designed carefully. Up to now, 2 LEHs cylindrical hohlraum is the main candidate. The laser beams are injected into the hohlraum through the 2 LEHs at both ends [3]. In this case, laser beams are arranged in multi rings to obtain the required symmetry. However, this may lead to crossed-beam energy transfer (CBET) [4-6], LPI problem of inner beams [3] and so on, which increase the difficulty of symmetry control. Rugby hohlraum [7-10] as another kind of 2 LEHs hohlraum has similar problems with cylindrical hohlraum. Besides 2 LEHs hohlraum, multiple-LEHs hohlraum has also been largely studied, such as 4 or 6 LEHs spherical hohlraum. Among them, multi rings are still necessary for symmetry control for the 4 LEHs spherical hohlraum [11-13]. And it may have the similar LPI problem as cylindrical hohlraum. 6 LEHs spherical hohlraum utilizes single-cone beams and has the natural superiority in radiation symmetry. In recent years, many theoretical and experimental studies on 6 LEHs spherical hohlraum have been carried out [14-18]. As an important alternative, a new kind of 6 LEHs hohlraum named TACH [19] has been proposed recently, which is jointed of three cylindrical hohlraums orthogonally. Laser beams are injected from 6 LEHs in single-cone with a large incident angle. Analysis shows that TACH has similar LPI as the outer ring of cylindrical hohlraum and similar symmetry as 6 LEHs spherical hohlraum. In addition, TACH has more flexibility on laser arrangement than 6 LEHs spherical hohlraum, which makes it possible to carry out high-quality integrated implosion at existing laser facilities.

This letter reports the first integrated implosion experiment of TACH. 24 laser beams of the SGIII laser facility [20,21] were carefully chosen and quasi-symmetrically injected into TACH. The radiation field inside TACH was studied by using array of flat x-ray detectors (FXRDs) [22] and shock wave technique (SWT). Backscatter fraction was studied by the Full Aperture Backscatter Station (FABS) and Near Backscatter Station (NBS). The implosion performance of a deuterium gas filled plastic capsule placed at the center of TACH was studied by diagnosing neutron yield, bang time, ion temperature and X-ray self-emission images of imploded capsule cores.

As shown in Fig.1, vacuum TACH was used in the experiment, which was orthogonally jointed of three cylindrical hohlraums. The length of each cylindrical hohlraum was 3.2mm, and the inner diameter was 1.4mm. A gold cube with a side length of 3.2mm was used to fabricate the hohlraum. As shown in Fig. 1, three orthogonal cylindrical holes with a diameter of 1.4mm were dug in the cube, and the inner surface of the three holes just formed the required TACH. For the SWT target, the witness material was inserted into the hohlraum by using a gold cone. For the implosion target, the cube was cut into two parts, and the capsule was placed at the center of the hohlraum by using support tent. The inner diameter of the capsule was 630μm. The ablator was silicon (Si) doped (1% by atom) CH, with a shell thickness of 35μm. The capsule was filled with deuterium gas at two different pressures, 13atm and 15atm.

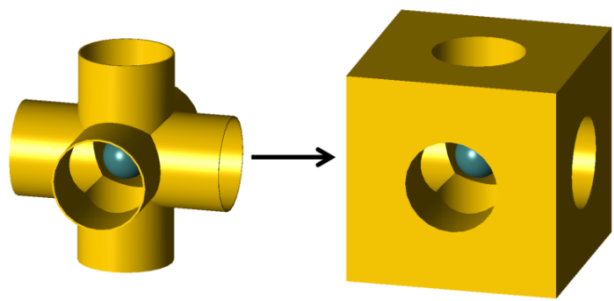

FIG. 1. Schematic of TACH. A gold cube was used to fabricate the TACH. Three orthogonal cylindrical holes with a diameter of 1.4mm were dug in the cube, and the inner surface of the three holes just formed the required TACH.

The experiment was performed at the SGIII laser facility, which consists of 48 laser beams at a wavelength of 0.35μm. There are four cones at the upper and lower hemisphere respectively [18]. To form a quasi-symmetric injection, 24 laser beams out of outer cones (cone3 and cone4) were carefully chosen, and each LEH was injected by 4 laser beams, as shown in fig. 2(a). The laser incident angles varied within 49.3±1.42°, so the 24 beams formed a 6 ends quasi-symmetric injection. A good radiation symmetry provided as below was obtained by further adjusting the position of laser spots inside the TACH. The drive flux symmetry on the capsule placed at the center of the TACH was studied by 3D view-factor code IRAD3D [23]. The spherical harmonic component $C_{lm}$ [19] was used to characterize the asymmetry. As the analysis in Ref. [19], $C_{40}$ and $C_{44}$ were the dominant asymmetry components. At the early stage, $C_{40}$=1.15%, $C_{44}$=1.34%, as shown in Fig. 2(b), and at the end of the 2.5ns laser pulse, $C_{40}$=1.35%, $C_{44}$=1.46%, as shown in Fig. 2(c). The other components was less than 0.5% during the whole laser pulse.

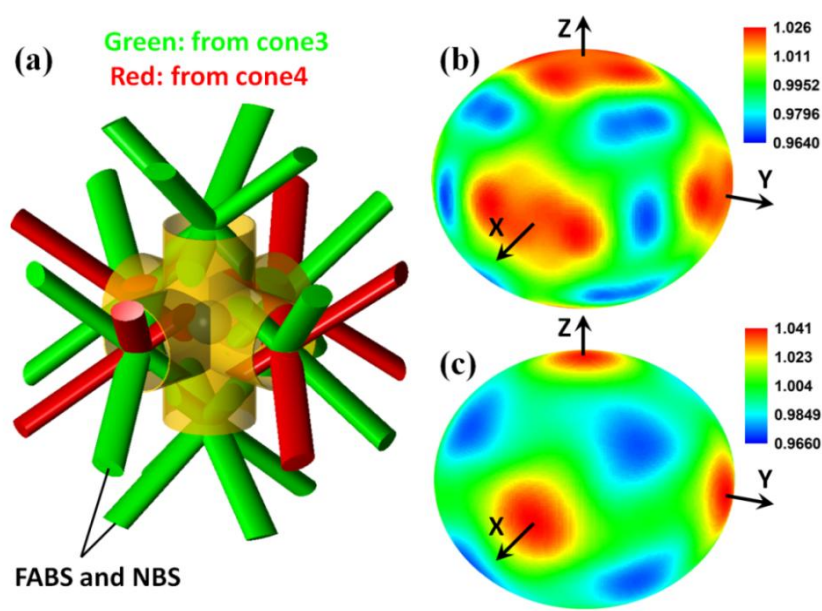

FIG. 2. (a). The 6 ends quasi-symmetric injection of TACH by use of 24 beams of SGIII laser facility, among which, the green and red beams belong to cone3 and cone4 respectively. (b). The drive distribution on the capsule surface at the early stage. (c). The drive distribution on the capsule surface at the end of the 2.5ns laser pulse. For (b) and (c), the drive distribution is normalized by the average drive flux, and the coordinate axes coincide with the three axes of TACH.

A 2.5ns square laser pulse was used in the experiments. The total energy delivered to the target was typically 46kJ. The 24 beams power balance ranged over 11%-15% rms. The rms beam pointing accuracy for the upper and lower LEHs was 30μm, while 80μm for the equatorial LEHs.

The radiation temperature was measured using array of FXRDs and SWT with Al step sample. 4 FXRDs (U16, D42, U42, U64) were used in the experiment, and were installed on different locations, and the corresponding view fields are shown in Fig.

3(a). Each FXRD can diagnose the radiation flux from several LEHs, and the measured flux is the sum of the fluxes from the LEHs in the view field. The equivalent radiation temperature of TACH can be obtained by use of the method described in Ref. [18], as shown in Eq. (1).

$$T_r = \left[ \frac{\pi F_{XRD}}{\sum_i^N \sigma A_{LEH}^{(i)} \cos\theta_i} \right]^{0.25} \quad (1)$$

The equivalent radiation temperature derived from Eq. (1) is shown in Fig. 3(b). The temporal behaviors of radiation temperature measured from different view angles are almost the same. The peak radiation temperature ranges from 186eV to 191eV with the variation being within the uncertainty of FXRD (3%), Which indicates that the radiation temperature measured by FXRDs is insensitive to observation location. For a given FXRD, when the location of which changes, its view angles from some LEHs become smaller, and from others become larger, the two conditions compensate each other, making the radiation temperature insensitive to the location of FXRD.

For the SWT target, Al step sample was placed near the inner surface of the TACH. X-ray radiation inside the hohlraum irradiated the sample, and generated a shock wave in the sample. The shock wave velocity was measured by Velocity Interferometer System for Any Reflector (VISAR) [24]. The typical experimental result is shown in Fig. 3(c). Assuming that the radiation flux irradiated on the Al sample has the similar temporal behavior as that measured by FXRD, the scaling formula between the peak radiation temperature and the shock wave velocity can be obtained by using the radiation hydrodynamic code HYADES [25], as shown in Eq. (2):

$$T_{r,eV} = 0.0109 V_{cm/s}^{0.6395} \quad (2)$$

where $T_r$ is the peak radiation temperature in unit of eV, and V is the shock wave velocity in unit of cm/s. The measured shock wave velocity was about 43km/s. According to Eq. (2), the peak radiation temperature irradiating the surface of Al sample is about 190eV, which agrees well with that measured by FXRD as shown in Fig. 3(d).

In Figure. 2(a), FABS and NBS were employed to measure the backscatter fraction of two laser beams from the lower LEH and one of the equatorial LEHs. The measured total backscatter fraction was less than 1.25%, and more than 2/3 of the backscatter energy was contributed by SBS, which was closed to the backscatter fraction of the outer ring laser beams in the vacuum cylindrical hohlraum at Omega laser facility [8,9]. The corresponding conversion efficiency of laser to X-ray is about 90%. It suggests that the laser behavior of TACH is similar to that of the outer ring of the traditional cylindrical hohlraum.

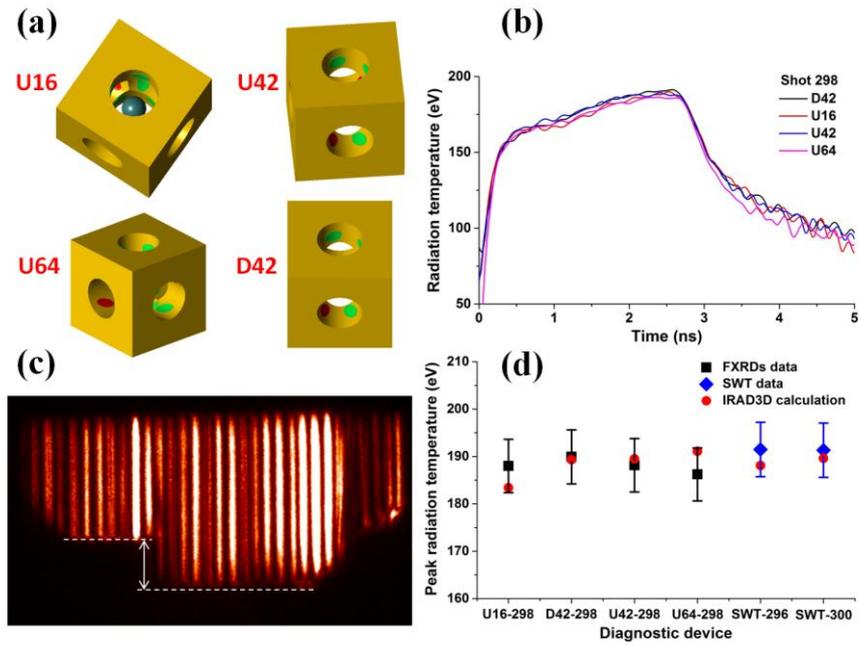

FIG. 3. (a) The view fields of the 4 FXRDs installed on the different locations of the target chamber. The green and red laser spots inside the hohlraum correspond to the cone3 and cone4 laser respectively. (b) The temporal radiation temperatures measured by the 4 FXRDs. (c) The typical VISAR streak image. (d) The comparison of peak radiation temperature between experiment and calculation, among which, the FXRD data (U16, D42, U42, U64) were measured in shot298, while the SWT data were measured in shot 296 and shot 300 respectively.

In the implosion experiment, the drive flux of the capsule could not be diagnosed directly. It should be associated with the flux measured by the FXRDs and SWT by use of numeric calculation. As shown in Fig. 3(d), the peak radiation temperatures calculated by IRAD3D agree well with that measured by FXRDs and SWT. So it is reasonable to calculate the drive flux of capsule through IRAD3D. The calculated peak radiation temperature on the surface of capsule is 185eV, and it is a little lower than that measured by FXRDs and SWT. This is mainly because the capsule can view large area of LEHs for the TACH used in this work, as shown in Fig. 1.

Different with traditional cylindrical hohlraum, six half cylindrical hohlraums of TACH were distributed symmetrically around the capsule. The gold bubbles created by the laser were separated from the capsule, which reduced the degradation effect of bubble plasma on the implosion performance of capsule. The drive symmetry was diagnosed from x-ray self-emission images of imploded capsule core by use of a time-integrated pin-hole camera through one of the equatorial LEHs. The bang time of neutron emission was measured by fast scintillator. The DD neutron yield was measured by scintillator. The ion temperature was measured by neutron time-of-flight (NTOF) device. The typical data are shown in Fig.4 and Table 1.

As shown in Fig. 4, a nearly round image of imploded capsule core was obtained. The symmetry of the core is analyzed from the 50% intensity contour of the image by use of spherical harmonic decomposition in sum of $\sum M_n * \exp[i * \sin(n\varphi)]$. The analysis gives $M_2/M_0=3.1\%$, $M_4/M_0=4.4\%$ and $M_6/M_0=2.9\%$. As described above, for the 6 ends quasi-symmetric injection, the dominant asymmetry component of the core is $M_4$, which is contributed by the $C_{44}$ drive asymmetry. The beam pointing accuracy for the upper and lower LEHs (30μm rms) is different from that for the equatorial LEHs (80μm rms), which may be the main cause for $M_2$. Besides that, the 24-beams power balance (11%-15% rms) and fabrication deviation of target could contribute to the $M_2$, $M_4$ and $M_6$ components. Even in the condition of imperfect beam pointing accuracy, power balance and non-ideal 6 ends injection, all of the spherical harmonic components are controlled below 5%, which presents the robustness of TACH in the symmetry control.

As shown in Table 1, DD neutron yield of $\sim 1.0 \times 10^9$ was obtained, and the corresponding Cr defined as the ratio of initial capsule outer radius to the hot spot radius at maximum neutron emission was 16~17. Higher Cr, ion temperature and DD neutron yield were obtained for lower gas pressure capsule. That was because lower gas pressure could achieve higher implosion velocity, which resulted in higher Cr and ion temperature. And higher ion temperature could increase fusion-reaction rate and compensate the lower fuel mass. So the DD neutron yield of 13atm capsule was yet higher than that of 15atm capsule. According to the radiation temperature studied above, clean-1D-neutron yield can be calculated. Then the YOC defined as the ratio of measured-neutron yield to clean-1D neutron yield can be obtained. In the experiment, a medium YOC of about 40% was obtained, which was obviously higher than that obtained in the 4 LEHs spherical hohlraum experiment at the same Cr regime [12]. In addition, some asymmetry factors may contribute to the degradation of YOC, such as the hydrodynamic mixing of cold ablator with the hot-spot region in the Cr regime, the preheating of capsule by hot electrons and M-band radiation generated by the interaction of laser with the hohlraum. Considering the non-ideal factors mentioned above and the asymmetry factors pointed out, the YOC obtained in the implosion experiment further reflects the robustness of radiation symmetry of TACH.

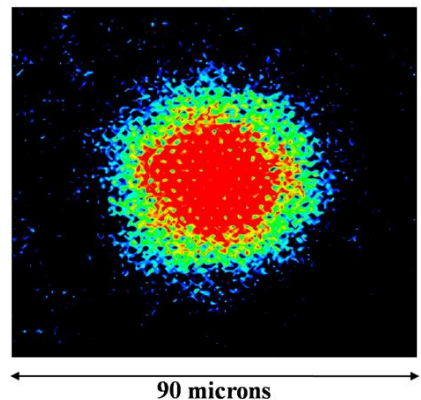

FIG. 4. The typical time-integrated X-ray self-emission images of imploded capsule core taken through one of the equatorial LEHs.

TABLE 1. The summary of implosion data

| Gas-filled pressure | 15 atm DD | 13 atm DD |
|---|---|---|
| Observed DD yield ($10^9$) | 1.0 | 1.1 |
| YOC | 39% | 41% |
| Calculate convergence ratio | 15.8 | 16.9 |
| Observed bang time (ns) | 2.7±0.2 | 2.7±0.2 |
| Observed ion temperature (keV) | 2.3 | 2.5 |

In conclusion, the integrated implosion experiment of TACH has been performed for the first time at the SGIII laser facility. 46kJ of frequency-tripled laser energy with 2.5ns square pulse was injected into TACH by a 6 ends quasi-symmetric injection arrangement. In the hohlraum energetics experiment, a quasi-symmetric radiation field with a peak radiation temperature of ~190eV was generated. The total backscatter fraction was less than 1.25%. In the implosion experiment, a nearly round core was observed and a DD neutron yield of ~$1 \times 10^9$ was obtained with YOC~40% and Cr~17. From the integrated implosion experiment, it was obvious that the LPI behavior of TACH was similar with that of the outer ring in traditional cylindrical hohlraum, and that the radiation field had the similar symmetry with that of the 6 LEHs Spherical Hohlraum. Therefore, TACH combines most advantages of various hohlraums and has little predictable risk, which provides an important potential way for ignition hohlraum design in ICF.

For further studies, to achieve a higher YOC with a higher Cr is more challenging and more relevant to ignition. It is crucial to evaluate the performance of TACH in the condition closer to ignition. To achieve this goal, it requires a comprehensive improvement in experimental technology and physical understanding.

The authors thank the SGIII operations crew, the target-fabrication groups and diagnose groups at Laser Fusion Research Center, whose efforts were crucial to the success of this work. This work was supported by the National Natural Science Foundation of China (Grand No. 11475154, 11505170, 11775204, 11435011).

[*]jiangshn@vip.sina.com

[†]ding-yk@vip.sina.com

[1] J. D. Lindl, Phys. Plasmas **2**, 3933 (1995).

[2] S. Atzeni and J. Meyer-ter-Vehn, *The physics of Inertial Fusion* (Oxford Science, Oxford, 2004).

[3] J. D. Moody *et al*., Phys. Plasmas **21**, 056317 (2014).

[4] P. Michel *et al.*, Phys. Rev. Lett. **102**, 025004 (2009).

[5] S. H. Glenzer *et al*., Science **327**, 1228 (2010).

[6] P. Michel *et al*., Phys. Rev. E **83**, 046409 (2011).

[7] M. Vandenboomgaerde *et al*., Phys. Rev. Lett. **99**, 065004 (2007).

[8] F. Philippe *et al*., Phys. Rev. Lett. **104**, 035004 (2010).

[9] H. F. Robey *et al.*, Phys. Plasmas **17**, 056313 (2010).
[10] P. E. Masson-Laborde *et al.*, Phys. Plasmas **23**, 022703 (2016).
[11] J. M. Wallace *et al.*, Phys. Rev. Lett. **82**, 3807 (1999).
[12] G. R. Bennett *et al.*, Phys. Plasmas **7**, 2594 (2000).
[13] S. Jiang *et al.*, Phys. Plasmas **23**, 122703 (2016).
[14] K. Lan *et al.*, Phys. Plasmas **21**, 010704 (2014).
[15] K. Lan *et al.*, Phys. Plasmas **21**, 052704 (2014).
[16] K. Lan *et al.*, Phys. Plasmas **21**, 090704 (2014).
[17] K. Lan *et al.*, Matter and Radiation at Extremes **1**, 8 (2016).
[18] W. Y. Huo *et al.*, Phys. Rev. Lett. **120**, 165001 (2018).
[19] L. Kuang *et al.*, Sci. Rep. **6**, 34636 (2016).
[20] W. Zheng *et al.*, High Power Laser Science and Engineering **4**, e21 (2016).
[21] W. Zheng *et al.*, Matter and Radiation at Extremes **2**, 243 (2017).
[22] Z. C. Li *et al.*, Rev. Sci. Instrum. **81**, 073504 (2010).
[23] L. Jing *et al.*, Phys. Plasmas **22**, 022709 (2015).
[24] L. M. Barker and R. E. Hollenbach, J. Appl. Phys. **43**, 4669 (1972).
[25] J. T. Larsen and S. M. Lane, J. Quant. Spectrosc. Radiat. Transf. **51**, 179 (1994).